# Mice Infected with Low-virulence Strains of *Toxoplasma gondii* Lose their Innate Aversion to Cat Urine, Even after Extensive Parasite Clearance

Wendy Marie Ingram[1*], Leeanne M Goodrich[1], Ellen A Robey[1, 3], Michael B Eisen[1, 2, 3]

**[1] Department of Molecular and Cell Biology, University of California, Berkeley, CA 94720, USA**
**[2] Howard Hughes Medical Institute**
**[3] Co-corresponding authors**
**\* E-mail: wingram@berkeley.edu**

## ABSTRACT

*Toxoplasma gondii* chronic infection in rodent secondary hosts has been reported to lead to a loss of innate, hard-wired fear toward cats, its primary host. However the generality of this response across *T. gondii* strains and the underlying mechanism for this pathogen-mediated behavioral change remain unknown. To begin exploring these questions, we evaluated the effects of infection with two previously uninvestigated isolates from the three major North American clonal lineages of *T. gondii*, Type III and an attenuated strain of Type I. Using an hour-long open field activity assay optimized for this purpose, we measured mouse aversion toward predator and non-predator urines. We show that loss of innate aversion of cat urine is a general trait caused by infection with any of the three major clonal lineages of parasite. Surprisingly, we found that infection with the attenuated Type I parasite results in sustained loss of aversion at times post infection when neither parasite nor ongoing brain inflammation were detectable. This suggests that *T. gondii*-mediated interruption of mouse innate aversion toward cat urine may occur during early acute infection in a permanent manner, not requiring persistence of parasite cysts or continuing brain inflammation.

## INTRODUCTION

*Toxoplasma gondii* is a highly prevalent and successful neurotropic protozoan parasite that infects mammals and birds and is found nearly everywhere in the world [1,2]. However, the parasite can only sexually reproduce in the feline gut, defining cats as the primary host [3]. When *T. gondii* infects an intermediate host such as rodents or humans, it infiltrates the central nervous system and forms slow-growing cysts inside neurons where it can persist for the life of the host [4]. In order to complete the parasite's life cycle, an infected secondary host must be eaten by a cat. Behavioral studies comparing infected and uninfected mice have suggested that rodents lose their innate, hard-wired fear of cat odors when chronically infected with *T. gondii* [5,6,7], presumably enhancing the transmission of the parasite to its primary host.

While intriguing, these studies had several limitations. Most *T. gondii* strains found in North America and Europe can be categorized into three well-defined clonal lineages called Type I, Type II, and Type III [8,9,10]. The majority of behavioral studies have used Type II strains, which are known to result in high parasite-cyst loads in the brains of mice and cause correspondingly high levels of immune-mediated brain inflammation [11,12,13]. This can result in general brain pathology causing extraneous motor and behavioral changes in infected mice, complicating the interpretation of predator aversion responses.

Type III parasites, in contrast, result in a lower cyst burden and cause less general brain inflammation [14,15]. Type I parasites are typically highly lethal in mice; however, the recent identification of *ROP5* genes as the mediators of acute virulence and the generation of attenuated parasites lacking this locus [16,17] permits long-term Type I infection studies in mice, perhaps due to a non-persistent infection following Immunity Related GTPase-mediated clearance. Extending mouse innate avoidance studies to include the less virulent Type III and attenuated Type I parasites described above could lead to clearer behavioral results, free of mitigating pathology-related changes.

**RESULTS AND DISCUSSION**

In order to evaluate the loss of aversion to cat urine in mice, we developed a rigorous high throughput behavioral assay. Mice were placed in a 15-inch by 7-inch enclosure with a small plastic dish affixed to one end. Either bobcat urine or rabbit urine was added to the dish and mice were allowed to explore the cage freely for one hour in the dark. The movements of each mouse were automatically recorded using Motor Monitor Smart Frames by Kinder Scientific, which are comprised of a grid of infrared beams and detectors. Each time the animal crosses a photobeam, the system records a 'beam break,' and provides the position and time spent in each designated area. The time spent close to the dish, defined as 'Near Target', or on the opposite end of the enclosure, the 'Avoidance Area', was assessed for up to 16 individually caged mice simultaneously (Figure 1A).

In the process of optimizing our behavioral assay, we confirmed the loss of aversion to cat urine in mice previously reported by others [6,18]. In preliminary experiments with male and female Balb/c mice, chronic infection with two Type II parasite strains (Pru and ME49) caused loss of innate aversion to bobcat urine (unpublished data). Due to slightly higher variability in female mouse behavior, likely due to higher levels of pathology [19], we performed all other experiments exclusively with males.

As anticipated, mice infected with Type II parasites succumbed to infection during both the acute and chronic phase. Surviving mice tended to have more complications associated with the progress of the infection as well as more 'sickness' behavior characterized in part by lower activity levels. Due to the variety of generalized pathology and since detailed studies of Type II infection as it relates to mouse behavior exist elsewhere, we did not continue investigations with Type II parasites.

To better characterize *T. gondii*'s ability to cause loss of innate aversion to cat urine in mice, we compared two additional parasite strains: an attenuated Type I and a low-virulence Type III. Male mice were assayed at three time points: three weeks, two months, and four months post infection.  Uninfected mice showed no place preference when exposed to non-predator control rabbit urine (Figure 1Bi), spending close to equal time in the Near Target Area and the Avoidance Area.  When exposed to bobcat urine, these same animals exhibited marked aversion, spending more time in the Avoidance Area than in the Near Target Area (Figure1Bii).  Type I- and Type III-infected animals behaved similarly to uninfected animals when exposed to rabbit urine, including total movements made, proportion of time spent in the periphery (unpublished data), and lack of aversion (Figure 1C).  Additionally, we performed a Hidden Cookie Test to evaluate general olfaction and observed no difference between uninfected and infected animals (Uninfected, Type I-, and Type III-infected animals found the cookie on average within $96\pm14$, $109\pm18$, and $123\pm31$ seconds, respectively where variance indicates Standard Error of the Mean). Infection with either attenuated Type I or Type III parasites resulted in complete loss of aversion to bobcat urine (Figure 1Biii and 1Biv).  This effect was observed at all three time points and appears to be an all-or-nothing effect (Figure 1C).  There were no 'non-responders' in either infection group, nor did the effect diminish with time.  These data show that the ability of *T. gondii* to disrupt innate predator aversion extends to all three major parasite clonotypes types.

We next investigated the parasite load and the immune response in the brains of mice infected with attenuated Type I and Type III parasites during chronic infection.  Following the final behavior experiment, we sacrificed the cohort over the course of 4 weeks.  We carefully harvested the brains and skull cap meninges from each animal and assessed 10% of homogenated brain and meninges for parasite load using semi-quantitative polymerase chain reaction (qPCR) targeting an abundant *T. gondii* gene family, B1 [20].  Type III-infected mice all had readily detectable parasite load, well above uninfected controls (Figure 2Ai).  In contrast, Type I-infected mice all had undetectable levels of brain-resident parasites (Figure 2Aii).

The brain is generally considered an immune-privileged environment and immune cell leukocyte infiltration is known to be tightly regulated [21,22].  Thus, only during an ongoing infection would we expect to find a large number of brain-resident leukocytes such as CD4 and CD8 positive T cells, required to control toxoplasmosis [23,24].  We isolated and compared brain leukocytes isolated from the combined parenchyma and meninges of each animal.  Purified cells were surface stained for a variety of leukocyte markers and analyzed using flow cytometry.  Type III-infected animals had elevated total brain leukocyte numbers, whereas attenuated Type I-infected animals had equivalent brain leukocyte numbers to uninfected animals (Figure 2B).  The average proportion of CD4 and CD8 positive T cells in Type III-, but not Type I-, infected animals were much higher than uninfected controls, reflecting ongoing brain inflammation. Brain leukocytes from uninfected controls and Type I-infected animals show a lower proportion of T cells, B cells, and neutrophils and likely correspond to meningeal leukocytes.  We assayed blood serum for antibodies specific for *T. gondii* using an enzyme-linked immunosorbent assay (ELISA) and confirmed that both attenuated Type I and Type III parasites had established infection in our mice (Figure 2C). Together these data show that permanent loss of aversion to predator urine may not depend on persistent brain infection.

It is possible that *T. gondii* causes a permanent change in the brain during acute infection, thereby not requiring continued parasite presence and an ongoing immune response. To address this possibility, we performed time course experiments, first infecting mice with Type III parasites. As expected, both parasite load and brain leukocyte numbers increased and remained well above uninfected levels (Figure 3A and 3B). We also infected mice with attenuated Type I parasites and detected parasites in the brain and meningeal homogenate in a number of mice between 5 and 20 days post infection (Figure 3D). At later time points, qPCR assays resulted in undetectable levels of these parasites. This early appearance of parasites in the brain may be related to the greater motility of Type I parasites in comparison to other strains [25], or may reflect more rapid dissemination within host cells. Early infection with attenuated Type I parasites was accompanied by a modest increase in total brain leukocytes (Figure 3E). While the Type I-infected animals had relatively low brain leukocyte numbers at day 13 post infection, the average percent of CD4 and CD8 positive T cells in these mice was similar to Type III-infected animals. Moreover, the increase in leukocyte infiltration is most striking when considering the total numbers of CD4 and CD8 T cells, providing further evidence for a T cell-mediated immune response in the brain and/or meninges following infection with attenuated Type I parasites (Figure 3F). Day 8 through day 20 post infection with attenuated Type I parasites from two separate experiments resulted in brain and meningeal T cell numbers all significantly above uninfected animals. This suggests that infection with attenuated Type I parasites does in fact elicit a transient inflammatory response in the brain and/or meninges following parasite presence.

Combined with previously published studies, our data indicate that infection with all three major North American *T. gondii* clonal lineages results in loss of innate, hard-wired aversion to feline predator urine in mice. Immunological analysis of mice infected with attenuated Type I and low-virulence Type III strains demonstrates that this behavioral change is not directly correlated with parasite load or brain inflammation. Taken together, our studies suggest that permanent interruption of mouse innate aversion to feline urine is a general trait of *T. gondii* infection that occurs within the first three weeks, independent of parasite persistence and ongoing brain inflammation.

Some current models propose that cysts residing in neurons play an active role in mediating loss of predator aversion. For example, it has been posited that *T. gondii* cysts might actively modulate dopamine production [26, 27, 28, 29], or directly interrupt neuronal activity [30, 31]. In line with this notion, some investigators report higher cyst density in amygdalar regions known to be involved in innate fear [6], although this has been challenged by others [32]. In contrast, our results suggest that cysts may not even be required for sustained fear disruption. Moreover, recent studies show that *T. gondii* can deliver effector proteins into cells that it does not invade [33, 34], and that these proteins can manipulate host cells without active parasite replication [35]. Thus *T. gondii* may interact with and manipulate its intermediate hosts without the requirement of cyst formation or parasite persistence. In light of these findings and our data reported here, we believe that a new non-cyst-centric model of *T. gondii*-mediated behavior manipulation of the mouse intermediate host is warranted.

## MATERIALS AND METHODS

### Ethics Statement

This study was carried out in strict accordance with the recommendations in the Guide for the Care and Use of Laboratory Animals of the National Institutes of Health. The protocol was approved by the Animal Care and Use Committee of UC Berkeley (Protocol #: R165-1212BCR).

### Animals

All mice were bred and housed in specific pathogen-free conditions at the Association of Laboratory Care-approved animal facility at the Life Science Addition, University of California, Berkeley, CA. Preliminary experiments used male and female BALB/c mice bred in our facilities ranging in age from 5 to 16 weeks old. Females were housed 5 animals per cage while males were housed between 1 and 5 animals per cage. Males were separated and housed individually if they began to fight on a cage-by-cage basis. Behavior experiments reported here and time course experiments involved male BALB/c mice all 9 weeks old, ordered from The Jackson Laboratory, 10 mice per group. Animals were housed 5 per cage until they began to fight. Upon the first cage needing to be separated, all animals were housed individually. Mice were sacrificed between 5 days and 20 weeks post infection by transcardial perfusion with 20 mL ice-cold sterile phosphate buffered saline following anesthesia with 500μL 2.5% avertin administered intraperitoneally (i.p.).

### Behavior Studies

Mouse aversion was assayed using the MotorMonitor SmartFrame System (Kinder Scientific, Poway, CA; Build # 11011-16). We used 7 x 15 High-Density SmartFrames to record up to 16 individually caged animals simultaneously. Animal movement was evaluated in transparent (17 cm X 38 cm) polycarbonate enclosures using the computerized photobeam system MotorMonitor. Animals were habituated for 1 hour the day before each trial in an empty enclosure. For the bobcat and rabbit urine exposure trials, a sterile cell culture dish (35mm x 10mm, treated polystyrene) was affixed to one end of the enclosure using a small amount of Blu-Tack (Bostick). 400 μL of either bobcat urine (LegUp Enterprises, Lovell, ME) or rabbit urine (Pete Rickard's, Fleming Outdoors, Ramer, AL) was added to the dish. One animal per cage was gently placed in the center of the enclosure and their activity monitored for 1 hour in the dark. The following day, we repeated the experiment with whichever urine sample was not used the previous day, the order of which was semi-randomized. Data was analyzed using MotorMonitor Software. User-defined Zone Maps were generated as described in Figure 1A. Recorded beam breaks were used by the software to quantify total movement and time spent in each zone. Heat maps were generated by MotorMonitor using the HotSpots Graphic Comparator for each animal trial (parameters: Time 1 hour, smoothing 0.1, Intensity Cube Rt, Minimum Visibility Normal). For the hidden cookie test, animals were food-deprived overnight (16-17 hours), during which time water was freely available. Testing performed on the following morning consisted of timing the latency for the animal to find an appetizing piece of food (1 Teddy Grahams cookie) buried 1 cm beneath fresh cage bedding.

### *T. gondii* infections

Tachyzoites were cultured on monolayers of human foreskin fibroblasts and prepared immediately before mouse infection as previously described [36]. All infections were administered i.p. in 200 μL volumes. Type I parasites used for infection were attenuated RH*Δrop5* (dose: $5x10^5$ parasites). Type II parasites used in preliminary studies were either Prugniaud expressing tandem dimeric tomato red fluorescent protein and OVA peptide (dose: 2500 parasites) or ME49 expressing luciferase (dose: 300 parasites). Type III parasites used for infection were CEP expressing green fluorescent protein (dose: $5x10^5$ parasites). All parasite strains were generously provided by John Boothroyd (Stanford, Palo Alto, CA).

### *Ex vivo* analysis of tissue samples

Brains and skull caps were harvested and placed in 10 mL ice cold PBS before being immediately processed. Brains were transferred to 2.5 mL cold RPMI medium 1640 + L-glutamine (Sigma) and meninges were dissected from the skullcaps as described elsewhere [37] and combined with the brain parenchyma. Brains and meninges were crushed using a 3 mL syringe plunger then homogenized by repeated passage through an 18-gauge needle. 10% of the homogenate was removed and stored at -20°C for future PCR analysis. The remaining homogenate was digested in 1 mg/mL Collagenase IA (Sigma) and 0.1 mg/mL DNase I (Roche) for 40 minutes at 37°C. The digested material was filtered through a 70 μm cell strainer and centrifuged at 800$g$ for 5 minutes. The brain and meningeal material was resuspended in room temperature 60% (vol/vol) Percoll (GE Healthcare)/RPMI and overlaid with 30% (vol/vol) Percoll/PBS and centrifuged with no acceleration or brakes for 20 minutes at 1000$g$. Mononuclear cells were harvested from the gradient interface and washed twice in PBS before preparation for flow cytometric analysis.

### Flow cytometry

Antibodies to mouse CD4 (RM4-5), CD8α (53-6.7), CD19 (eBioID3), CD11b (M1/70), and CD11c (N418) were obtained from eBioscience. Anti-mouse Ly6G (1A8) was obtained from BD Biosciences. Cell viability was assayed using LIVE/DEAD® Fixable Aqua Dead Cell Stain Kits (Invitrogen). Surface staining with anti-mouse CD4, CD8, CD19, CD11b, CD11c, and Ly6G was performed at 4°C for 30 minutes. Cells were fixed and acquisitions were performed using a BD LSR II flow cytometer (BD) and data were analyzed with FlowJo software (Tree Star, Ashland, OR).

### Parasite load assay

Whole genome DNA was isolated from brain and meningeal homogenate using DNeasy Blood and Tissue Kit (Qiagen). Parasite burden was assessed using semi-quantitative PCR as described elsewhere [38].

**Serum ELISA for *T. gondii***

Blood was collected by mandibular vein bleed prior to animal sacrifice. Samples were incubated for 4 hours at room temperature to allow clot to form, then incubated at 4°C overnight. ELISA 96 well plates were coated with 1 μg/ml Soluble Tachyzoite Antigen (STAg). Plates were washed 3 times with PBS – 0.05% TWEEN and blocked with 5% milk PBS-0.05% TWEEN, then washed 3 times more with PBS-0.05% TWEEN, and 54 μl of 5% milk PBS-0.05% TWEEN was applied to each well. Five serial dilutions, 6μl each, of serum from each sample were incubated overnight at 4°C. Plates were washed three times with PBS-0.05% TWEEN, and once with PBS. Plates were then incubated with rabbit anti-mouse IgG (H+L)-HRP (Jackson Immuno Research) 1:1000 dilution for 2 hours at room temperature. $H_2O_2$ was added to ABTS substrate (Sigma) at 1:1000, mixed and applied to the plate. Enzymatic reaction times were recorded at 405 nm.

**Statistical Analysis**

Prism software (GraphPad) was used for statistical analysis. P values were calculated using two-tailed Student's (non-parametric) t-test or 1 way ANOVA as indicated.

**ACKNOWLEDGEMENTS**

The authors would like to gratefully acknowledge John Boothroyd and Michael Reese (Stanford) for their generous gift of the attenuated Type I RH*Δrop5* parasite, Shiao Chan and Kayleigh Taylor for technical assistance, and members of the E.A.R. and M.B.E. labs for helpful comments, support, and assistance.

**FIGURE LEGENDS**

**Figure 1. Assessment of aversion demonstrates loss of fear toward cat urine in Type I- and Type III-infected mice.** [A] Overhead representation of behavioral arena where a small dish containing the 'target' solution (yellow disk) is affixed at one end of the behavioral arena. 'Near Target' is defined as the area of the arena (white) proximal to the target. 'Avoidance' is defined as the most distal region (dark grey) of the enclosure relative to the target. [B] Representative heat maps of mouse place preference during a 60-minute trial of (i) uninfected mice exposed to rabbit urine, and (ii) uninfected, (iii) attenuated Type I-infected, and (iv) low-virulence Type III-infected mice exposed to bobcat urine from trials conducted at 2 months post infection. [C] Aversion ratio, the avoidance time to near target time, of uninfected (red circles), Type I-infected (green triangles), and Type III-infected (blue squares) animals when exposed to bobcat urine (filled shapes) or rabbit urine (open shapes) at 3 weeks, 2 months, and 4 months post infection (n=10 for each group). Error bars are the Standard Error of the Mean (SEM).

**Figure 2. Persistent inflammation during chronic infection with Type III but not Type I parasites.** [A] Quantitative PCR of genomic DNA prepared from 10% of brain homogenate reveals that parasite DNA was readily detectable in Type III-infected animals (i) but undetectable in attenuated Type I-infected animals (ii). [B] Brain leukocytes were percoll purified from the remaining brain and meningeal homogenate, stained for surface markers CD4, CD8, CD19, CD11b, and Ly6G, and assayed using flow cytometry. Total numbers of cells positive for any of these markers are reported for uninfected, Type I-infected, and Type III-infected animals at 4 to 5 mpi. Average percentages of brain leukocyte populations for uninfected, Type I, and Type III animals are displayed in pie charts below each group. [C] Blood serum was collected following the final behavior experiment 4 months post infection (4 mpi). *T. gondii* specific antibodies were detected using ELISA. Relative absorbance at 405 nm (sample 1:10 dilution absorbance – HRP no serum control absorbance) is reported for uninfected, Type I, and Type III. Each dot signifies one mouse. Significance was determined by Student's T-test for [A] and 1 way ANOVA for [B] and [C] where ns indicates p>0.05 and **** indicates p<0.0001 (n=9 for each group).

**Figure 3. Type I and Type III acute infection results in parasite and leukocyte infiltration of the brain region.** [A and D] Parasite presence was assessed using semi-quantitative PCR of genomic DNA prepared from 10% of each mouse brain and meninges at various times post infection. For reference, data from Figure 2A is included which was collected from animals that were used in behavior experiments 4 to 5 months post infection. [A] Type III-infected animals have detectable parasite in brain regions

early during acute infection, which is sustained over time. [D] Some attenuated Type I-infected animals have detectable parasite in brain regions early during acute infection, decreasing to undetectable levels (average of uninfected indicated by black dashed line). [B and E] Brain leukocytes were percoll purified from the remaining brain and meningeal homogenate, stained for leukocyte surface markers CD4, CD8, CD19, CD11b, and Ly6G, and assayed using flow cytometry. [B] Type III-infected animals have brain leukocyte numbers above uninfected levels (black dashed line) 13 days following infection which continue to increase over time. [E] Average total cell numbers from Type I-infected animals. Average percentages of brain leukocyte populations for uninfected, Type I-, and Type III-infected animals at selected time points are represented in pie charts below the corresponding data in [B and E]. [C and F] Total brain and meningeal T cells (CD4 plus CD8), indicators of brain region inflammation, are reported for Type III- and attenuated Type I-infected animals. Each dot signifies one mouse. Significance was determined for [C and F] by Student's T-test where * indicates $p < 0.02$, and ** indicates $p < 0.002$.

**Figure 1**

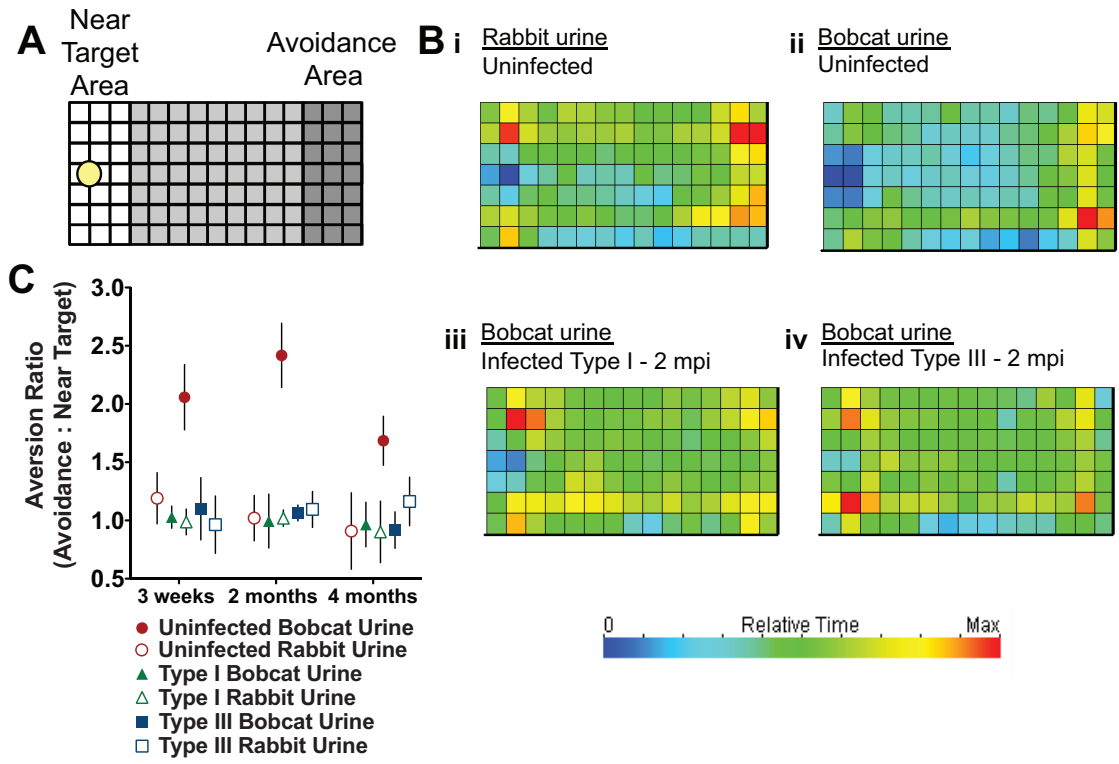

**A** Near Target Area    Avoidance Area

**B  i** Rabbit urine
Uninfected

**ii** Bobcat urine
Uninfected

**iii** Bobcat urine
Infected Type I - 2 mpi

**iv** Bobcat urine
Infected Type III - 2 mpi

**C**

Aversion Ratio
(Avoidance : Near Target)

3 weeks    2 months    4 months

● Uninfected Bobcat Urine
○ Uninfected Rabbit Urine
▲ Type I Bobcat Urine
△ Type I Rabbit Urine
■ Type III Bobcat Urine
□ Type III Rabbit Urine

0    Relative Time    Max

## Figure 2

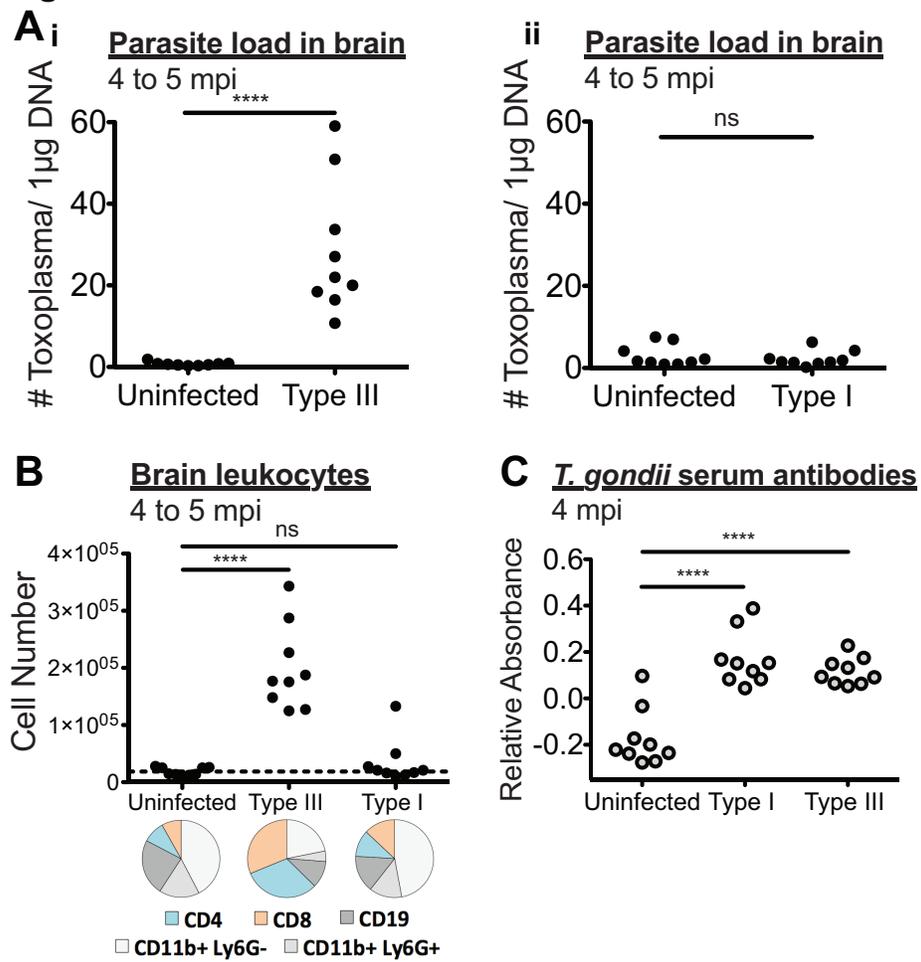

**A** i **Parasite load in brain**
4 to 5 mpi

ii **Parasite load in brain**
4 to 5 mpi

**B** **Brain leukocytes**
4 to 5 mpi

**C** *T. gondii* serum antibodies
4 mpi

CD4  CD8  CD19
CD11b+ Ly6G-  CD11b+ Ly6G+

**Figure 3**

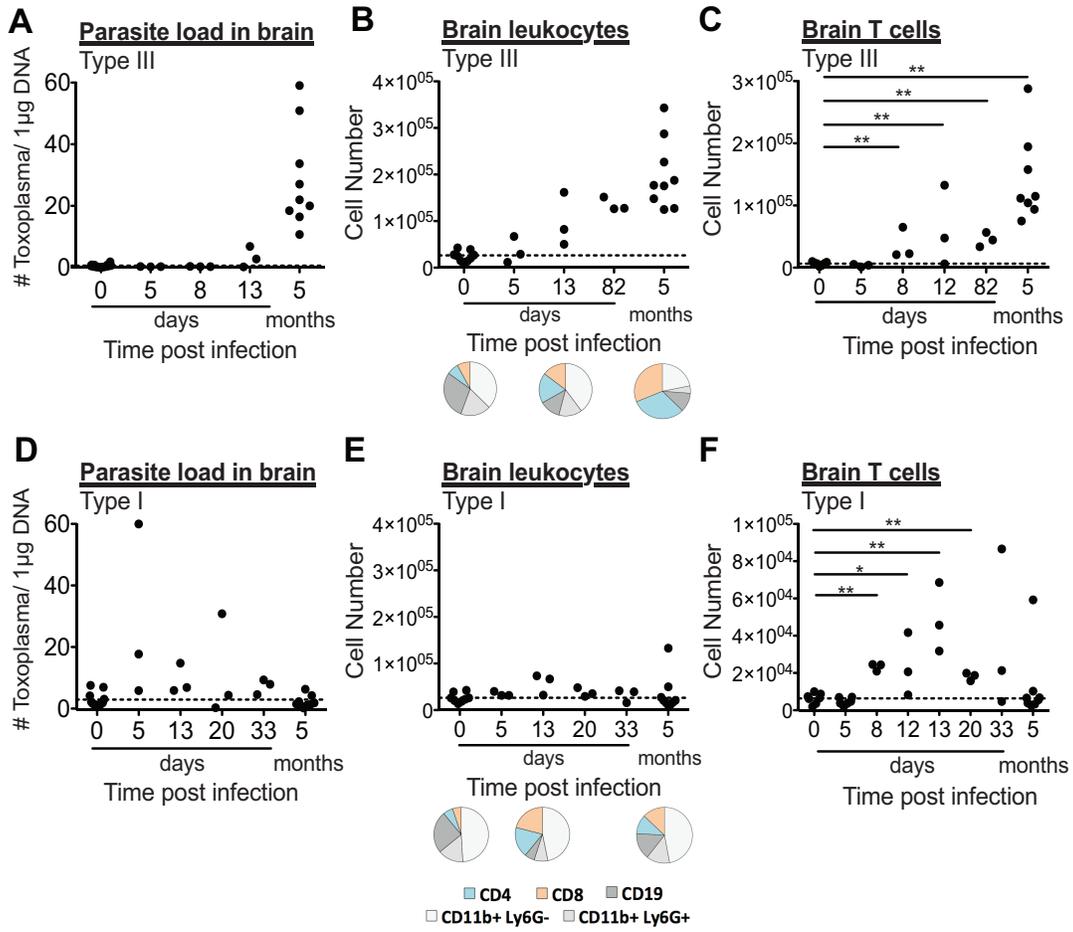